\documentclass[structabstract]{aa}  
\newcommand{\unit}[1]{\ensuremath{\, \mathrm{#1}}}
\usepackage{url}

\usepackage{graphicx}
\usepackage{natbib}
\usepackage[english]{babel}
\bibpunct{(}{)}{;}{a}{}{,}
\newcommand{\teff}{$T_{\rm eff}$}
\newcommand{\logteff}{$\log(T_{\rm eff})$}
\newcommand{\logg}{$\log g$}
\newcommand{\cd}{d$^{-1}$}
\newcommand{\GD}{$\gamma$~Dor}
\newcommand{\DSct}{$\delta$~Sct}
\newcommand{\vsini}{$v\sin{i}$}
\newcommand{\kms}{km\,s$^{-1}$}
\usepackage{txfonts}
\begin{document}
   \title{KIC 11285625: a double-lined spectroscopic binary with a $\gamma$ Dor
pulsator discovered from \textit{Kepler} space photometry.\thanks{Partly based
on observations made with the Mercator Telescope, operated on the island of La
Palma by the Flemish Community, at the Spanish Observatorio del Roque de los
Muchachos of the Instituto de Astrofísica de Canarias. 
Based on observations obtained with the HERMES spectrograph, which is supported
by the Fund for Scientific Research of Flanders (FWO), Belgium , the Research
Council of K.U.Leuven, Belgium, the Fonds National Recherches Scientific (FNRS),
Belgium, the Royal Observatory of Belgium, the Observatoire de Gen\`{e}ve,
Switzerland and the Th\"{u}ringer Landessternwarte Tautenburg, Germany.}}

   \author{
    J. Debosscher \inst{1}
   \and
     C. Aerts\inst{1,2}
   \and
     A. Tkachenko\inst{1}
   \and
     K. Pavlovski\inst{3}
   \and 
     C. Maceroni\inst{4}
   \and 
     D. Kurtz \inst{5} 
   \and  
     P. \,G.~Beck\inst{1}
   \and
     S. Bloemen\inst{1}
   \and
     P. Degroote\inst{1,7}
   \and
     R. Lombaert\inst{1}
   \and 
     J. Southworth\inst{6}
}
   \institute{Instituut voor Sterrenkunde, KU Leuven, Celestijnenlaan 200B, 3001
Leuven, Belgium
           \and
          Department of Astrophysics, Radboud University Nijmegen, POBox 9010,
6500 GL Nijmegen, the Netherlands
          \and 
Department of Physics, Faculty of Science, University of Zagreb, Croatia
          \and INAF - Osservatorio Astronomico di Roma via Frascati 33, 00040
Monteporzio C. (RM), Italy
          \and Jeremiah Horrocks Institute, University of
Central Lancashire, Preston PR1 2HE, UK
          \and Astrophysics Group, Keele University, Staffordshire ST5 5BG, UK
          \and Stellar Astrophysics Center, Department of Physics and Astronomy,
Aarhus University, Ny Munkegade 120, 8000 Aarhus C, Denmark 
        }

   \date{}

 
  \abstract
   {}
   {We present the first binary modelling results for the pulsating eclipsing
binary KIC 11285625, discovered by the \textit{Kepler} mission. An automated
method to disentangle the pulsation spectrum and the orbital variability in
high quality light curves, was developed and applied. The goal was to obtain
accurate orbital and component properties, in combination with essential
information derived from spectroscopy.}
   {A binary model for KIC 11285625 was obtained, using a combined analysis of
high-quality space-based \textit{Kepler} light curves and ground-based high-resolution HERMES echelle spectra. The binary model was used to separate the
pulsation characteristics from the orbital variability in the \textit{Kepler}
light curve in an iterative way. We used an automated procedure to perform this
task, based on the JKTEBOP binary modelling code, and adapted codes
for frequency analysis and prewhitening of periodic signals. Using
a disentangling technique applied to the composite HERMES spectra, we obtained a higher
signal-to-noise mean component spectrum for both the primary and the secondary. A model grid search method for fitting
synthetic spectra was used for fundamental parameter determination for both components.}
   {Accurate orbital and component properties of KIC
11285625 were derived, and we have obtained the pulsation spectrum of
the
$\gamma$ Dor pulsator in the system. Detailed analysis of the
pulsation spectrum revealed amplitude modulation on a time scale of a hundred
days, and strong indications of frequency splittings at both the orbital
frequency, and the rotational frequency derived from spectroscopy.}
   {}

  \keywords{Stars: oscillations; binaries: eclipsing; binaries: spectroscopic;
techniques: photometric; stars: fundamental parameters; stars: rotation}
  \titlerunning{KIC 11285625: a double-lined spectroscopic binary with a
$\gamma$ Dor pulsator}
   \maketitle
%

\section{Introduction}
NASA's \textit{Kepler} mission has been continuously monitoring more than
150\,000 stars for the past 4 years, searching for transiting exoplanets
\citep{Borucki-Kepler}.
The unprecedented quality of the photometric light curves delivered by
\textit{Kepler} makes them also very well suited to study stellar variability in
general. Automated light curve classification techniques with the goal to
recognize 
and identify the many variable stars hidden in the \textit{Kepler} database have
been developed. The application of these methods to the
public \textit{Kepler} Q1 data is described in \cite{Debosscher:2011}. There,
the authors paid special attention to the detection of pulsating stars 
in eclipsing binary systems. These systems are relatively rare, and especially
interesting for asteroseismic studies \citep[see
e.g.][]{Maceroni:2009,Welsh:2011}. By modelling the orbital dynamics of the
binary, using photometric time series complemented with spectroscopic follow-up
observations, we can
obtain accurate constraints on the masses and radii of the pulsating stars.
These
constraints are needed for asteroseismic modelling, and are difficult to obtain
otherwise. 

Numerous candidate pulsating binaries were identified in the
\textit{Kepler} data, and spectroscopic follow-up is ongoing. In this work, we
present the results obtained for KIC 11285625 (BD+48 2812), which turns out to be an
eclipsing binary system containing a $\gamma$ Dor pulsator. The KIC (\textit{Kepler} Input Catalog) lists the following properties for this target: $V = 10.143$ \,mag,  \teff = 6882 K, \logg = 3.753,
$R$ = 2.61 $\unit{R_{\sun}}$ and $[Fe/H]$ = -0.127.
Spectroscopic follow-up revealed it to be a double-lined binary (Section \ref{rv}). Currently, only a few $\gamma$ Dor pulsators in double-lined spectroscopic binaries are known, making their analysis very relevant for asteroseismology. \cite{Maceroni:2013} studied a $\gamma$ Dor pulsators in an eccentric binary system, observed by CoRoT. Here, we are dealing with a non-eccentric system with a longer orbital period. The longer time span and the higher photometric precision of the \textit{Kepler} observations (almost a factor 6) allowed us to study the pulsation spectrum with significantly increased frequency resolution and down to lower amplitudes.

A combined
analysis of the \textit{Kepler} light curve and spectroscopic radial velocities
allowed us to obtain a good binary model for KIC 11285625, resulting in accurate
estimates of the masses and radii of both components (Section \ref{binmod}). 
This binary model was also
used to disentangle the pulsations from the orbital variability in the
\textit{Kepler} light curve in an iterative way. In this paper, we describe
procedures to
perform this task in an automated way. The low signal-to-noise composite
spectra used for the determination of the radial velocities were used to obtain
higher signal-to-noise mean spectra of the components, by means of
spectral disentangling. These spectra were then used to obtain fundamental
parameters of the stars (Section \ref{disentangling}). 
The resulting pulsation signal of the primary is analysed in detail in Section \ref{puls-spec}.
There we discuss the global characteristics of the frequency spectrum, we list
the dominant frequencies and their amplitudes detected by means of prewhitening,
and search for signs of rotational splitting.

\section{\textit{Kepler} data}

\begin{figure*}
 \centering
  
\includegraphics[width=14cm,angle=270,scale=0.7]{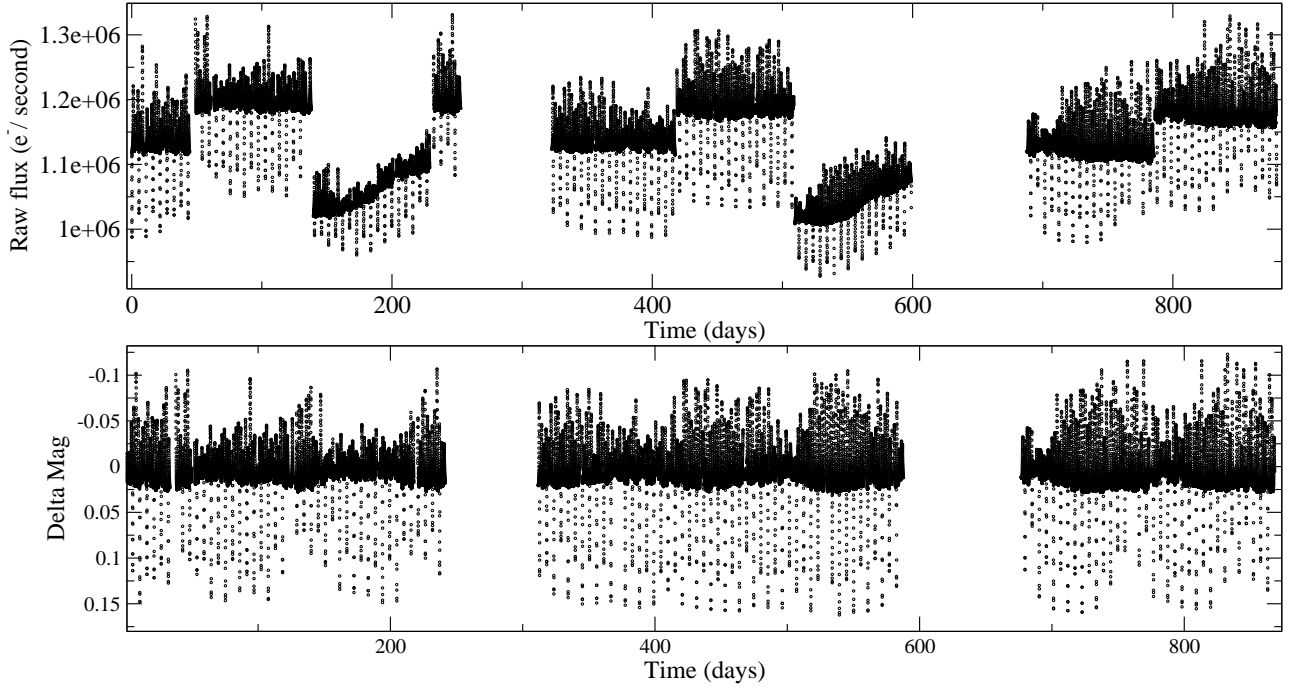}
\caption{Combination of observing quarters Q1-Q10 for KIC 11285625. The top
panel shows the `raw'
SAP (simple aperture photometry) fluxes, as they were delivered, while the lower
panel shows the resulting combined dataset after optimal mask selection and detrending.}
\label{LC-all-quarters}
\end{figure*}

KIC 11285625 has been almost continuously observed by \textit{Kepler}; data are
available for observing quarters Q0-Q10. We only used long cadence data in this work (with a time resolution of 29.4 minutes), since short cadence data is only available for three quarters and is not needed for our purposes. During quarter Q4, one of the CCD modules failed, the reason why part of the Q4 data are missing. No Q8 data could be observed either, since
the target was positioned on the same broken CCD module during that quarter.
The \textit{Kepler} spacecraft needs to make rolls every 3 months (for
continuous illumination of its solar arrays), causing targets to fall on
different CCD modules depending on the observing quarter. 
Given the different nature of the CCDs, and the different aperture masks used,
this caused some issues with the data reduction. The average flux level of the
light curve for KIC 11285625 varies significantly 
between quarters, and for some, instrumental trends are visible. The top
panel of Fig. \ref{LC-all-quarters} plots all the observed
datasets, showing the quarter-to-quarter
variations. Merging the quarters correctly is not
trivial, since the trends have to be removed for each quarter separately, and 
the data have to be shifted so that all quarters are at the same average level
(see below).
Often, polynomials are used to remove the trends, but it is difficult to
determine a reasonable order for the polynomial. This is especially the case
when large amplitude variability, at time scales comparable to the total time
span of the data, is present in the light curve.

\begin{figure*}
 \centering
   \includegraphics[width=14cm,angle=0,scale=1.0]{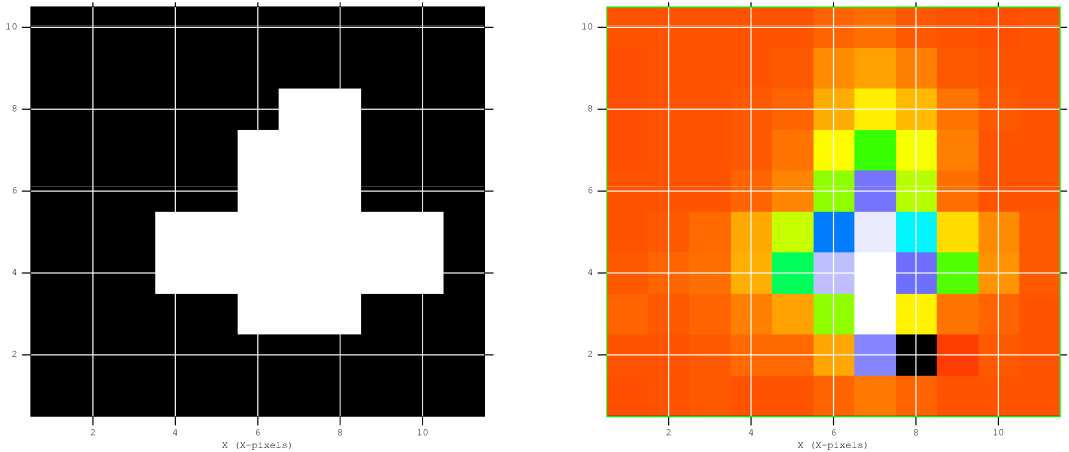}
\caption{\textit{Kepler} aperture (left) and a single target pixel image (right)
for KIC 11285625, quarter Q3. Red colours in the pixel image indicate the lowest flux levels, white colours indicate the highest flux levels.}
\label{mask}
\end{figure*}

Fortunately, pixel target files are available for all observed quarters for
KIC 11285625, allowing us to do the light curve extraction based on custom aperture masks. We can define a custom aperture mask, determining
which pixels to include or not. It turned out that the standard aperture mask, used by
the data reduction pipeline, was not optimal for all quarters. 
This is clearly visible in the top panel of Fig. \ref{LC-all-quarters} for
quarters Q3 and Q7. The clear upward trends and smaller variability amplitudes
compared to the other quarters are caused by a suboptimal aperture mask. The
trends can be explained by a small drift of the star on the CCD, changing the
amount of stellar flux included in the aperture during the quarter.
Checking the target pixel files for those quarters revealed that pixels with
significant flux contribution were not included in the mask. Fig. \ref{mask}
shows the \textit{Kepler} aperture from the automated pipeline, and a single
target pixel image obtained during quarter Q3. As can be seen, the
\textit{Kepler} aperture misses a pixel
with significant flux contribution (the lowest blue coloured pixel).   
Adding this pixel to the aperture mask effectively removed the trend in the
light curve and increased the variability amplitude to the level of the other
quarters.

An automated method was developed to optimize the aperture mask for each
quarter, with the goal to maximize the signal-to-noise ratio (S/N) in the
Fourier amplitude spectrum. 
The standard mask provided with the target pixel files is used as a starting
point. The
method then loops over each pixel in the images outside of the original mask
delivered by the \textit{Kepler} pipeline. For each of those pixels, a new light
curve is constructed by adding the flux values of the pixel
to the summed flux of the pixels within the original \textit{Kepler} mask. The
amplitude spectrum of the resulting light curve is then computed and the S/N of
the highest peak (in this case, the main pulsation frequency of the star) is
determined.
If the addition of the pixel increases the S/N (by a user specified amount), it
will be added to the final new light curve once each pixel has been analysed
this way. The method also avoids adding pixels containing significant flux of
neighbouring contaminating targets, since these will normally decrease
the S/N of the signal coming from the main target. It is also possible to detect
contaminating pixels within the original \textit{Kepler} mask, using exactly the
same method, but now by excluding one pixel at a time from the original mask 
and checking the resulting S/N of the new light curves.

After determining a new optimal aperture mask for each quarter, the resulting
light curves still showed some small trends and offsets, but they were easily
corrected using second order polynomials.  
Special care is needed however when shifting quarters to the same level, since
the average value of the light curve might be ill determined, especially
when large amplitude non-sinusoidal variability is present at time scales
similar
to the duration of an observing quarter.
In our case, we want to make the average out-of-eclipse brightness match
between different quarters, and not the global average of the quarters, since
the latter is shifted due to the presence of the eclipses. 
Therefore, we cut out the eclipses first and interpolated the data points in the
resulting gaps using cubic splines. This was done only to determine the
polynomial coefficients, which were then used to subtract 
the trends from the original light curves (including the eclipses). We first
transformed the fluxes into magnitudes for each quarter separately, prior to
trend removal. The binary modelling code we describe further needs magnitudes
as input, but the conversion to magnitude also
resolves any potential quarter-to-quarter variability amplitude changes caused, e.g., by a difference in CCD gains (or any other instrumental effect changing the
flux values in a linear way).        

The lower panel of Fig. \ref{LC-all-quarters} shows the resulting light curve
after application of our detrending procedure using pixel target files.

\section{Spectroscopic follow-up and RV determination}
\label{rv}
Spectroscopic follow-up observations were obtained with the HERMES Echelle
spectrograph at the Mercator telescope on La Palma (see \cite{HERMES} for a detailed
description of the instrument). 
In total, 63 spectra with good orbital phase coverage were observed, with
S/N in \textit{V} in the range 40-70. These spectra revealed the
double-lined nature of the spectroscopic binary. 

Radial velocities were derived with the HERMES reduction pipeline, using the
cross-correlation technique 
with an F0 spectral mask. The choice for this mask was based on the F0 spectral
type given by SIMBAD and the effective temperature listed in the KIC (Kepler
Input Catalogue). We also tried additional spectral masks, given the
double-lined
nature of the binary, but we did not obtain better results in terms of scatter
on the radial velocity points. Due to the numerous metal lines in the spectrum,
the cross-correlation technique worked very well, despite the relatively low S/N
spectra. The upper panel of  Fig. \ref{rv-lc} shows the radial velocity measurements obtained for
both components of KIC 11285625.
The black circles correspond to the primary and the red circles to the
secondary. From the scatter on the radial velocity measurements, we conclude
that the primary component is pulsating. A Keplerian model was fitted for both
components (shown also in the upper panel of Fig. \ref{rv-lc} ), resulting in the orbital parameters listed in Table \ref{final-par}.

Uncertainties were estimated using a Monte-Carlo perturbation approach. 
Although the orbital period of the system can be a free parameter in the fitting
procedure for the Keplerian model, we fixed it to the much more accurate value obtained from the
\textit{Kepler} light curve, given its longer time span (see Section \ref{binmod}). The quality
of the fit obtained this way (as judged from the $\chi^2$ values) is significantly better compared to the case where
the orbital period is left as a free parameter.

\section{Binary model}
   
\label{binmod}
We used the combined \textit{Kepler} Q1-Q10 data to obtain a binary model,
which, combined with the results from the spectroscopic analysis, provided us
with accurate estimates of the main astrophysical properties
of both components. Given that we are dealing with a detached binary with no or only
limited distortion of both components, we used JKTEBOP, written by
J. Southworth \citep[see][]{Southworth:JKTEBOP1,Southworth:JKTEBOP2}.
This code is based on the EBOP code, originally developed by Paul B. Etzel
\citep[see][]{Etzel:EBOP,Popper:EBOP}. JKTEBOP has the advantage of being very
stable, fast, and it is applicable to large datasets with thousands of
measurements, such as the \textit{Kepler} light curves. 
Moreover, it can easily be scripted (essential in our approach) and includes
useful error analysis options such as Monte Carlo and bootstrapping methods.\\

\begin{figure*}
 \centering
  
\includegraphics[width=14cm,angle=270,scale=0.7]{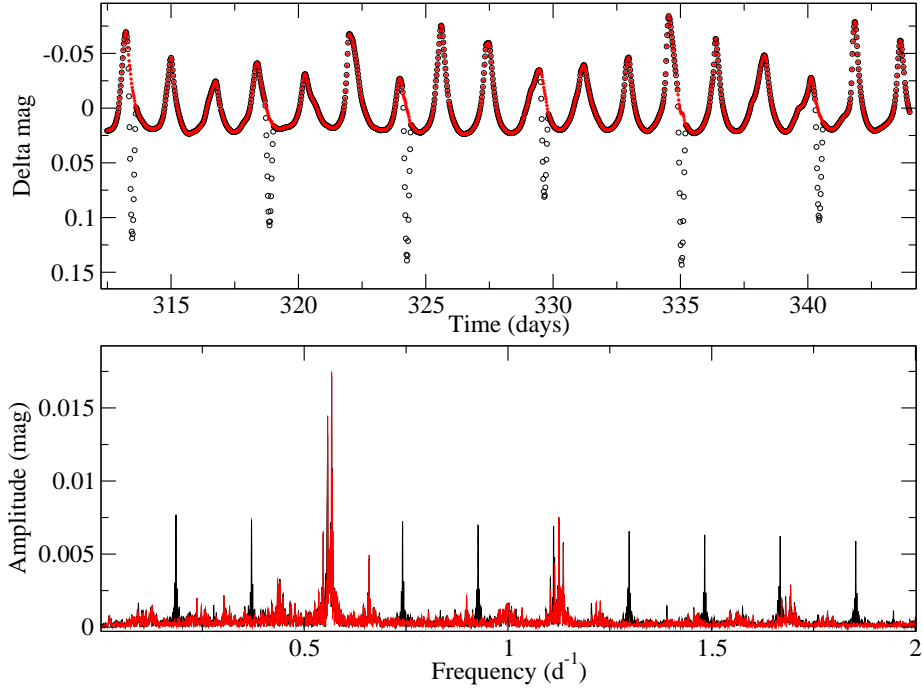}  
\caption{Upper panel: resulting light curve after iterative removal of the
orbital variability (in red), with the original light curve shown for comparison
(in black). Lower panel: Amplitude spectrum of the original
\textit{Kepler} light curve (in black), and the 
amplitude spectrum of the pulsation `residuals' (in red) after iterative
removal of the orbital variability.}
\label{puls-residuals}
\end{figure*}

Since we are dealing with a light curve containing both orbital variability
(eclipses) and pulsations, we had to disentangle both phenomena in order to
obtain a reliable binary model. 
In the amplitude spectrum of the \textit{Kepler} light curve (see Fig.
\ref{puls-residuals}), the $\gamma$ Dor type pulsations have numerous significant peaks in
the range 0-0.7 $\unit{d^{-1}}$, with clear repeating patterns up to around 4 $\unit{d^{-1}}$. In the same region of the amplitude spectrum, we
also find peaks 
corresponding to the orbital variability: a comb-like pattern of harmonics of
the orbital frequency ($f_\mathrm{orb}$, $2f_\mathrm{orb}$, $3f_\mathrm{orb}$,...). Moreover, the
main pulsation frequency of 0.567 $\unit{d^{-1}}$ is close to $6f_\mathrm{orb}$ (0.556
$\unit{d^{-1}}$ ), though the peaks are clearly separated, given an estimated
frequency resolution
of $ 1/T \approx0.0012\, d^{-1}$, with T the total time span 
of the combined \textit{Kepler} Q1-Q10 data. This near coincidence complicated the disentangling
of both types of variability in the light curve, unlike the case where the
orbital variability is well separated from the 
pulsations in frequency domain (e.g. typical for a $\delta$ Sct pulsator in a
long-period binary). We used an iterative procedure, consisting of an
alternation
of binary modelling with JKTEBOP, and prewhitening of 
the remaining variability (pulsations) after removal of the binary model. This
procedure can be done in an automated way, and the number of iterations can be
chosen. The idea is that we gradually improve both the binary model and the
residual pulsation spectrum at the same time. In each step of the procedure, we
used the entire Q1-Q10 dataset without any rebinning. 
Our iterative method is similar to the one described in \cite{Maceroni:2013} and
consists of the following steps: 
\begin{enumerate}
 \item Remove the eclipses from the original \textit{Kepler} light curve and
interpolate the resulting gaps using cubic splines.
 \item Derive a first estimate of the pulsation spectrum by means of iterative
prewhitening of the light curve without eclipses.
 \item Remove the pulsation model derived in the previous step from the original
\textit{Kepler} light curve.
 \item Find the best fitting binary model to the residuals using JKTEBOP.
 \item Remove this binary model from the original \textit{Kepler} light curve
(dividing by the model when working in flux, or subtracting the model when
working in magnitudes).
 \item Perform frequency analysis on the residuals, which delivers a new
estimate of the pulsation spectrum.  
 \item Subtract the pulsation model obtained in the previous step from the
original \textit{Kepler} light curve. 
 \item Model the residuals (an improved estimate of the orbital variability)
with JKTEBOP and repeat the procedure starting from step five. 
\end{enumerate}

\begin{figure}
 \centering
   \includegraphics[width=14cm,angle=0,scale=0.66]{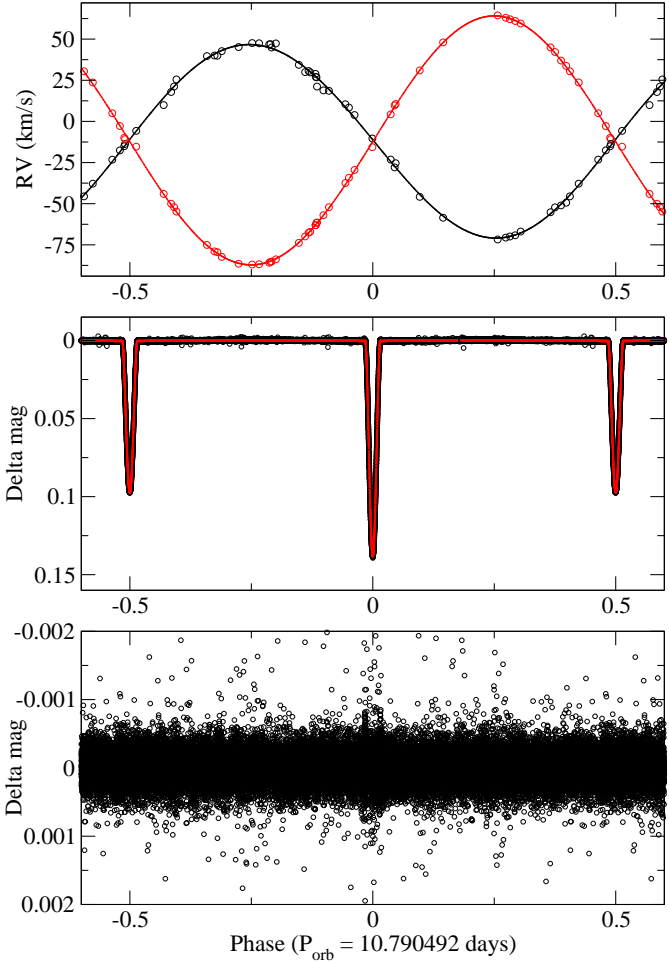}
\caption{Upper panel: phased HERMES radial velocity data (open circles) and Keplerian model fits (lines) for both components (black: primary, red: secondary). Middle panel: phased \textit{Kepler} data (pulsations removed) with the binary model overplotted in red.
Lower panel: phased residuals of the \textit{Kepler} data, after removal of both the pulsations and the binary model. }
\label{rv-lc}
\end{figure}

The procedure is then stopped when convergence is obtained: the $\chi^{2}$
value of the binary model no longer decreases significantly.
In practice, convergence is obtained after just a few iterations, at least for
KIC 11285625. The procedure can be run automatically,
provided that the user has a good initial guess for the orbital parameters.
The computation time is dominated by the prewhitening
step, which requires the repeated calculation of amplitude spectra. In our case,
the complete procedure required a few hours on a single desktop CPU. The top panel of Fig. \ref{puls-residuals} shows part of the original \textit{Kepler} data,
with the disentangled pulsation contribution overplotted in red, the lower
panel shows the corresponding amplitude spectrum. In Fig.
\ref{rv-lc}, the Keplerian model fit to the HERMES RV data, the binary model
fit to the \textit{Kepler} data (pulsation part removed) and the residual light
curve are shown together, phased with the orbital period (using the
zero-point $T_\mathrm{0}$ = 2454953.751335 d, corresponding to a time of minimum
of the primary eclipse). The slightly larger scatter in the residuals at the
ingress
and egress of the primary eclipse is caused by the fact that the
occulted surface of the pulsating primary is non-uniform and changing over
time, due to the non-radial pulsations.

We also investigated a different iterative approach, where we started by fitting
a binary model directly to the original \textit{Kepler} light curve, instead of
first removing the pulsations. This way, we do not remove information from the
light curve by cutting the eclipses, and we do not need to interpolate the data.
Although the iterative procedure also converged quickly using this approach, the
final binary model was not accurate. Removing the model from the original
\textit{Kepler} light curve introduced systematic offsets during the eclipses.
The reason is that the initial binary model obtained from the original light
curve is inaccurate due to the large amplitude and non-sinusoidal nature of the
pulsations, causing the mean light level between eclipses to be badly defined. 
The approach starting from the light curve with the pulsations removed prior to
binary modelling provided much better results, as judged from the
$\chi^{2}$ values and visual inspection of the 
residuals during the eclipses.\\

A linear limb-darkening law was used for both stars, with coefficients
obtained from \cite{Prsa:2011}\footnote{\url{http://astro4.ast.villanova.edu/aprsa/?q=node/8}}. 
These coefficients have been computed for a grid of {\teff},
{\logg} and [M/H] values, taking into account \textit{Kepler's} transmission,
CCD quantum efficiency and optics. We estimated them using the KIC
(Kepler Input Catalogue) parameters for a first
iteration, but later adjusted them using our obtained values for {\teff},
{\logg} and [M/H] from the combination of binary modelling and spectroscopic
analysis (see Section \ref{disentangling}). We did not take gravity darkening and refection effects into account, given that both components are well separated, and are not significantly deformed by rapid rotation or binarity (the oblateness values returned by JKTEBOP are very small).

During the iterative procedure, special care was paid to the following
issues, since they all influence the quality of the final binary and
pulsation models:

\begin{itemize}
 \item Removal of the pulsations from the original light curve: here, we first
determined all the significant pulsation frequencies (or, more general:
frequencies most likely not caused by the orbital motion) from the light curve
with the
eclipses removed. Only frequencies with an amplitude signal-to-noise ratio (S/N)
above four are considered. The noise level in the amplitude spectrum is
determined from the region 20- 24 d$^{-1}$, where no significant peaks are
present. The often used procedure of computing the noise level in a region
around the peak of interest would not provide reliable S/N estimates in our
case, given the high density of significant peaks at low frequencies. 
We also used false-alarm probabilities as an additional significance test,
with very similar results regarding the number of significant frequencies. 

\item Initial parameters of the binary model: when running JKTEBOP, initial
parameters have to be provided for the binary model. These are then refined
using non-linear optimization techniques (e.g. Levenberg-Marquardt). 
Although the optimization procedure is stable and converges fast, we have no
guarantee that the global best solution is obtained. If the initial parameters
are too far off from their true values, the procedure can end up in a local
minimum. Therefore, we did some exploratory analysis first, to find a good set
of initial 
parameters, aided by the constraints obtained from the radial velocity data. The
initial parameters were also refined: after completion of the first
iterative procedure, the final parameters were used as initial values for a new
run, etc. This also confirmed the stability of the solution, although it does
not guarantee that the overall best solution has been found.  
\end{itemize}

Error analysis of the final binary model was done using the Monte-Carlo method
implemented in JKTEBOP. Here, the input light curve (with the pulsations
removed) is perturbed by adding Gaussian noise with standard deviation estimated
from the residuals, and the binary model is recomputed. This procedure is
repeated typically 10000 times, to obtain confidence intervals for the obtained
parameter values.
The final parameters from the combined spectroscopic and photometric analysis,
and their estimated uncertainties (1$\sigma$), are listed in Table \ref{final-par}.

Given the long time span and excellent time sampling of the light curve, we also checked for the presence of eclipse time variations (e.g. due to the presence of a third body).
We used two different methods to check for deviations of pure periodicity of the eclipse times. The first method consists of computing the amplitude spectrum of each observing quarter 
of the light curve separately, and comparing the peaks caused by the binary signal (the comb of harmonics of the orbital frequency). We could not detect any change in orbital frequency this way.
Moreover, the orbital peaks in the amplitude spectrum of the entire light curve also do not show any broadening or significant sidelobes (indicative of frequency changes), compared to the amplitude spectrum of the purely periodic light curve of our best binary model computed with JKTEBOP. The second method consists of determining the times of minima for each eclipse individually by fitting a parabola to the bottom of each eclipse and determining the minimum. We then compared those times with the predicted values using our best value of the orbital period (O-C diagram). This was done both for the original light curve and the light curve with pulsations removed. In the first case, we found indications of periodic shifts of the eclipse times, but these are clearly linked to the pulsation signal in the light curve, which also affects the eclipses. No periodic shifts or trends could be found when analysing the light curve with pulsations removed, and we conclude that we do not detect eclipse time variations.

\begin{table}
\tiny
\renewcommand{\tabcolsep}{0.4mm}
\center
\caption{Orbital and physical parameters for both components of KIC 11285625,
obtained from the combined spectroscopic and photometric analysis.}
 \begin{tabular}{lcc}
 \hline
 &System&\\
 \hline
 Orbital period  $P$ (days)&  10.790492 $\pm$ 0.000003  &\\
 Eccentricity $e$ &$0.005\pm0.003$ &\\
 Longitude of periastron $\omega$ (\degr)&90.103 $\pm$ 0.006&\\
 Inclination $i$ (\degr) & 85.32 $\pm$ 0.02 &\\
 Semi-major axis $a$ ($\unit{R_{\sun}}$) &28.8 $\pm$ 0.1&\\
 Light ratio $L_1/L_2$ &0.38 $\pm$ 0.01&\\
 System RV $\gamma$ (km $\unit{s^{-1}}$)&-11.7 $\pm$ 0.2&\\
 $T_\mathrm{0}$ (days)\tablefootmark{a}&2454953.751335 $\pm$ 0.000014 \\
 \hline  
 &Primary&Secondary\\
 \hline
 Mass $M$ ($\unit{M_{\sun}}$)&$1.543\pm0.013$ &$1.200\pm0.016$\\
 Radius $R$ ($\unit{R_{\sun}}$) & 2.123 $\pm$ 0.010& 1.472 $\pm$ 0.014\\
 log $g$ &3.973 $\pm$ 0.006 & 4.18 $\pm$ 0.01 \\
\hline
\end{tabular}
\tablefoottext{a}{Reference time of minimum of a primary eclipse.}
\label{final-par}
\end{table}

\section{Spectral disentangling}
\label{disentangling}

To derive the fundamental parameters  \teff, \logg, $[M/H]$, etc. of the $\gamma$ Dor pulsator from
the high
resolution HERMES spectra, we first needed to separate the contributions
 of both stars in the measured composite spectra. Given the similar spectral types of the components, and the large number of metal lines in the spectra, 
we could apply the technique of spectral disentangling to accomplish this
\citep{Simon:1994,Hadrava:1995}. Here,
we used the FDBinary code \footnote{http://sail.zpf.fer.hr/fdbinary/}
\citep{Ilijic:2004} which is based on Hadrava's Fourier approach
\citep{Hadrava:1995}. The overall procedures used to determine the orbital
parameters and reconstruct the spectra of the component stars from time series
of observed composite spectra of a spectroscopic double-lined eclipsing binary
have been described extensively in \cite{Hensberge:2000} and
\cite{Pavlovski:2005}. 

The user has to provide good initial guesses and confidence intervals for the
orbital parameters, otherwise the method might not converge towards the correct
solution. Luckily, we had very good
initial parameter values available from the Keplerian orbital fit to the radial
velocity data, as was described in Section \ref{rv}. The final orbital
parameters
turned out to be in excellent agreement with the initial values derived from
spectroscopy. 

\begin{figure}

 \centering

\includegraphics[width=14cm,angle=270,scale=0.5]{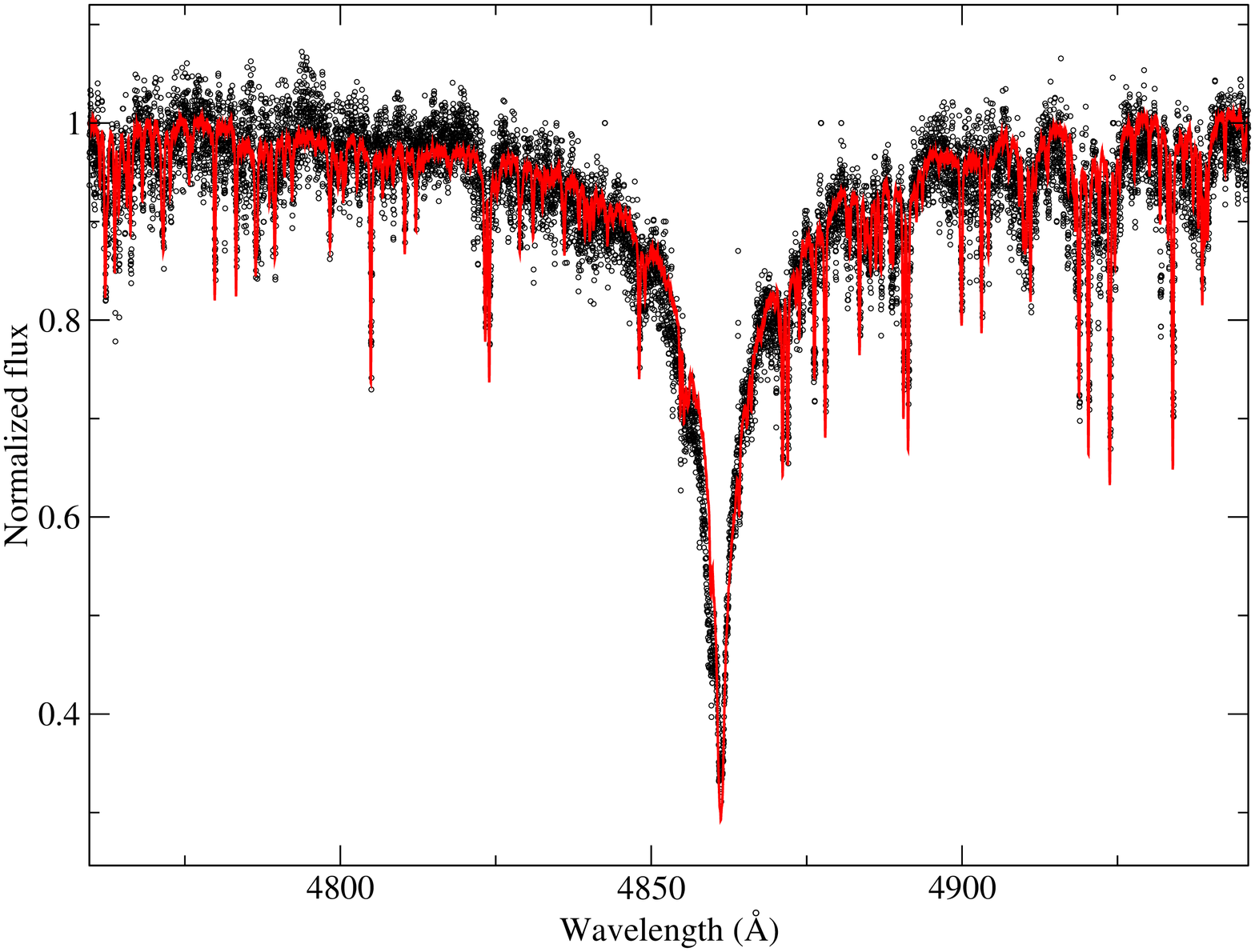}
\caption{Comparison of a single observed composite spectrum (black circles) with
the disentangled spectrum of the primary component (red lines), illustrating
the significant increase in S/N that is obtained.}
\label{disentangled-primary}
\end{figure}

Spectral disentangling methods have the advantage that they enable us to
determine
the orbital parameters of the system in an independent way (although good
initial estimates are necessary), and that the resulting component spectra have
a higher signal-to-noise ratio than the individual original composite spectra:
S/N $\sim$ $\sqrt{N}$, with N the number of composite spectra used for
disentangling. The increase in S/N is illustrated in
Fig. \ref{disentangled-primary}, where a single observed spectrum (corrected for
Doppler shift) is compared to the disentangled spectrum for the primary
component. Renormalization of the disentangled spectra was done using
the light factors obtained from the binary modelling of the \textit{Kepler}
light
curve. 

For the spectrum analysis of both components of KIC\,11285625, we
use the GSSP code \citep[Grid Search in Stellar Parameters,][]{Tkachenko2012} that finds the optimum
values of \teff, \logg, $\xi$, $[M/H]$, and \vsini\ from the minimum
in $\chi^2$ obtained from a comparison of the observed spectrum with
the synthetic ones computed from all possible combinations of the
above mentioned parameters. The errors of measurement (1$\sigma$
confidence level) are calculated from the $\chi^2$ statistics, using the
projections of the hypersurface of the $\chi^2$ from all grid points
of all parameters in question. In this way, the estimated error bars include
any possible model-inherent correlations between the parameters but do not take
into account imperfections of the model (such as incorrect atomic data, non-LTE
effects, etc.) and/or continuum normalization. A
detailed description of the method and its application to the
spectra of  \textit{Kepler} $\beta$\,Cep and SPB candidate stars as well as
$\delta$\,Sct and $\gamma$\,Dor candidate stars are given in
\citet{Lehmann2011} and \citet{Tkachenko2012}, respectively.

For the calculation of synthetic spectra, we used the LTE-based
code SynthV \citep{Tsymbal1996} which allows the computation of the
spectra based on individual elemental abundances. The code uses
calculated atmosphere models which have been computed with the
most recent, parallelised version of the LLmodels program
\citep{Shulyak2004}. Both programs make use of the VALD database
\citep{Kupka2000} for a selection of atomic spectral lines.
The main limitation of the LLmodels code is that the models are
well suited for early and intermediate spectral type stars, but
not for very hot and cool stars where non-LTE effects or
absorption in molecular bands may become relevant, respectively.

\begin{table}
\caption{Fundamental parameters of both components of KIC\,11285625.}\label{Table: Fundamental parameters}
\begin{tabular}{lll}
\hline\hline
\multicolumn{1}{c}{Parameter\rule{0pt}{9pt}} & Primary & \multicolumn{1}{c}{Secondary}\\
\hline
\teff\,(K)\rule{0pt}{11pt} &$6960\pm 100$ & $7195\pm200$\\
\logg\,(fixed)\rule{0pt}{11pt} & 3.97 & 4.18\\
$\xi$\,(\kms)\rule{0pt}{11pt} & $0.95\pm0.30$ & $0.09\pm0.25$\\
\vsini\,(\kms)\rule{0pt}{11pt} & $14.2\pm1.5$ & $8.4\pm1.5$\\
$[M/H]$\,(dex)\rule{0pt}{11pt} & $-0.49\pm0.15$ & $-0.37\pm0.3$\\
\hline
\end{tabular}
\tablefoot{The temperature of the secondary is not reliable, as discussed in the text.}
\end{table}

Given that KIC\,11285625 is an eclipsing, double-lined (SB2)
spectroscopic binary for which unprecedented quality (\textit{Kepler})
photometry is available, the masses and the radii of both components
were determined with very high precision. Having those two
parameters, we evaluated surface gravities of the two stars with
far better precision than one would expect from the spectroscopic
analysis given that the S/N of our spectra varies between 40 and 70,
depending on the weather conditions on the night when the
observations were taken. Thus, we fixed \logg\ for both
components to their photometric values (3.97 and 4.18 for the
primary and secondary, respectively) and adjusted the effective
temperature \teff, micro-turbulent velocity $\xi$, projected
rotational velocity \vsini, and overall metallicity [M/H] for both
stars based on their disentangled spectra.
Given that the contribution of the primary component to the total
light of the system is significantly larger than that of the
secondary (72\% compared to 28\%) and that its decomposed spectrum
is consequently better defined and is of higher quality than that of
the secondary, we were also able to evaluate individual abundances
for this star besides the fundamental atmospheric parameters.
Table~\ref{Table: Fundamental parameters} lists the fundamental
parameters of the two stars whereas Table~\ref {Table: Individual abundances} summarizes the results
of chemical composition analysis for the primary component. The overall metallicities of the two
stars agree within the quoted errors, but the derived temperature for the
secondary is not reliable, since we find 
it to be about 200~K hotter than the primary. From the relative eclipse depths,
we estimate the temperature of the secondary to be about 6400~K. The cause of
this temperature discrepancy is the poor quality of the disentangled spectrum of
the secondary, given its smaller light contribution. Normalization errors in
the spectra can easily translate into temperature errors of several hundred
Kelvin. Note that the listed uncertainties for the spectroscopic temperatures do not take normalization errors into account.

Figure~\ref{Figure: HR diagram} shows the position of the primary component in
the \logteff-\logg\ diagram, with respect to the observational  \DSct\  (solid
lines) and \GD\ (dashed lines) instability strips as given by
\cite{Rodriguez:2001} and \cite{Handler:DSCUT}, respectively. The primary falls
into the \GD\ instability strip, meaning that pure g-modes are expected to be
excited in its interior.

\begin{table}
\tabcolsep 2.2mm\caption{Atmospheric chemical composition of the
primary component of KIC\,11285625.}
\label{Table: Individual abundances}
\begin{tabular}{llllll}
\hline\hline
\multicolumn{1}{c}{Element\rule{0pt}{9pt}} & Value & \multicolumn{1}{c}{Sun} & \multicolumn{1}{c}{Element\rule{0pt}{9pt}} & Value & \multicolumn{1}{c}{Sun}\\
\hline
Fe\rule{0pt}{11pt} & --0.58\,(15) & --4.59 & Mg & --0.65\,(23) & --4.51\\
Ti\rule{0pt}{11pt} & --0.40\,(20) & --7.14 & Ni & --0.53\,(20) & --5.81\\
Ca\rule{0pt}{11pt} & --0.41\,(27) & --5.73 & Cr & --0.35\,(27) & --6.40\\
Mn\rule{0pt}{11pt} & --0.50\,(35) & --6.65 & C & --0.28\,(40) & --3.65\\
Sc\rule{0pt}{11pt} & --0.31\,(40) & --8.99 & Si & --0.68\,(40) & --4.53\\
Y\rule{0pt}{11pt} & --0.22\,(45) & --9.83 &  &  & \\
\hline
\end{tabular}
\tablefoot{All values are in dex and on a
relative scale (compared to the Sun). Label ``Sun'' refers to the
solar composition given by \citet{Grevesse2007}. Error bars
(1-$\sigma$ level) are given in parentheses in terms of last digits.}
\end{table}

\begin{figure}
\includegraphics[scale=0.9,clip=]{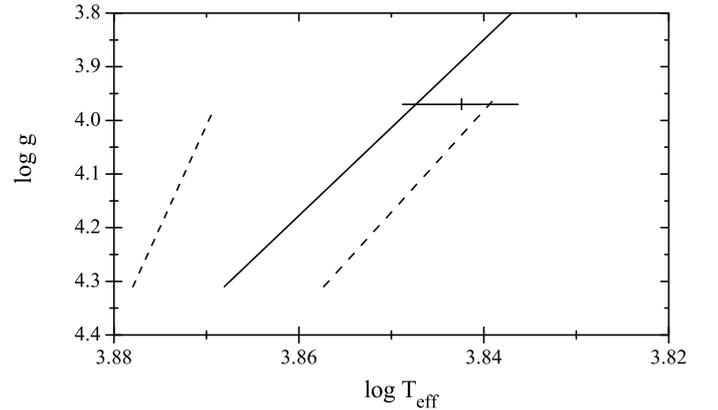}
\caption{{\small Location of the primary of KIC\,11285625 in the
\logteff-\logg\ diagram. The \GD\ and the red edge of the \DSct\
observational instability strips
are represented by the dashed and solid lines, correspondingly. According to
the photometric \teff\ and \logg, the secondary would be located in the lower
right corner of the diagram, outside both instability strips. }} \label{Figure:
HR diagram}
\end{figure}

\section{Pulsation spectrum}
\label{puls-spec}

\begin{figure*}
 \centering
   \includegraphics[width=14cm,angle=270,scale=0.7]{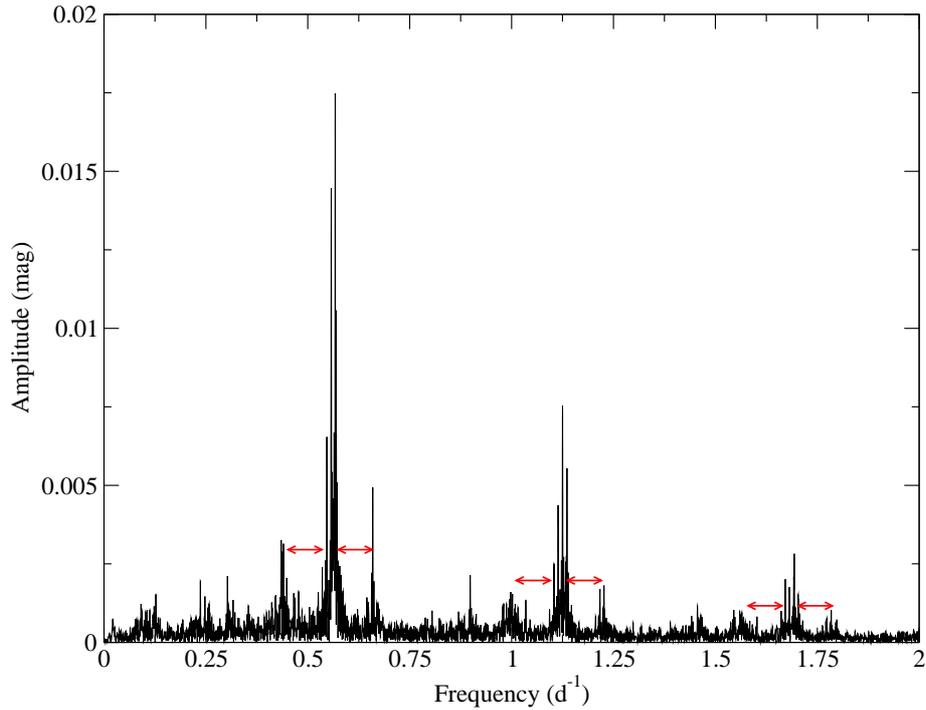}
\caption{Part of the amplitude spectrum (below 2 {\cd}) of the \textit{Kepler}
light curve with the binary
model removed. The red arrows indicate the splitting of groups of peaks caused
by rotation with a  splitting value around 0.1 {\cd}.}
\label{as-puls-global}
\end{figure*}

Fig. \ref{as-puls-global} shows the amplitude spectrum of the \textit{Kepler}
light curve with the binary model removed, in the region 0-2 d$^{-1}$, where
most of the dominant pulsation frequencies are found. Clearly visible
are the three groups of peaks around 0.557, 1.124 and 1.684 {\cd}. Some of the
frequencies found in the groups around 1.124 and
1.684 {\cd} are harmonics of frequencies present in the group around
0.557 {\cd}. These harmonics are caused by the
non-linear nature of the pulsations, and have been observed for many pulsators
observed by CoRoT and  \textit{Kepler}, along almost the entire main-sequence
\citep{Degroote:2009,Poretti:2011,Breger:2011,Balona:2012}, and in \GD\ stars in
particular \citep{Tkachenko:2013}. Closer
inspection of the
main frequency groups revealed that they consist of several closely spaced
peaks,
almost
equally spaced with a frequency of $\sim$ 0.010 {\cd}. A likely explanation is
amplitude modulation of the pulsation signal on a timescale of $\sim$ 100
days. Mathematically, the effect of amplitude modulation can be
described as follows: imagine a simplified case where a single periodic
non-linear
pulsation signal is being modulated with a general periodic function. We can
write this signal as a product of two sums of sines, where the number of terms
(harmonics) in each sum depends on how non-linear (non-sinusoidal) the signals
are: 
\begin{equation}
\sum_{i=1}^{N_\mathrm{mod}} a_\mathrm{i} \sin{[2\pi f_\mathrm{mod}
i t + \phi^\mathrm{mod}_\mathrm{i}]} \sum_{j=1}^{N_\mathrm{puls}} b_\mathrm{j} \sin{[2\pi f_\mathrm{puls}
j t + \phi^\mathrm{puls}_\mathrm{j}]} ,
\end{equation}
with $f_\mathrm{mod}$ the modulation frequency and $f_\mathrm{puls}$ the pulsation frequency.
This product can be rewritten as a sum, using Simpson’s rule:
\begin{eqnarray}
\sum_{i=1}^{N_\mathrm{mod}} \sum_{j=1}^{N_\mathrm{puls}} \frac{a_\mathrm{i} b_\mathrm{j}}{2}(\cos{[2\pi
(f_\mathrm{puls} j-f_\mathrm{mod} i)t+\phi^\mathrm{puls}_\mathrm{j}-\phi^\mathrm{mod}_\mathrm{i}]}-\\
\cos{[2\pi (f_\mathrm{puls} j+f_\mathrm{mod} i)t+\phi^\mathrm{puls}_\mathrm{j}+\phi^\mathrm{mod}_\mathrm{i}]}). \nonumber
\end{eqnarray}
In the Fourier transform of this signal, we will thus see peaks at
frequencies which are linear combinations of the pulsation frequency and the
modulation frequency, where the number of combinations depends on how
non-linear
both signals are. Most of the observed substructure in the amplitude spectrum
can be explained by amplitude modulation of a pulsation signal with
$f_\mathrm{mod}\sim$ 0.010 {\cd}. A more detailed description of amplitude
and frequency modulation in light curves can be found in \cite{Benko:2011}.
Since the period of the amplitude modulation is close to the length of a
\textit{Kepler} observing quarter, we checked for a possible instrumental
origin. A very strong argument against this, is the fact that the modulation is
not present in the eclipse signal of the light curve, but is only affecting the
pulsation peaks. 

\begin{figure}
 \centering
   \includegraphics[width=14cm,angle=270,scale=0.5]{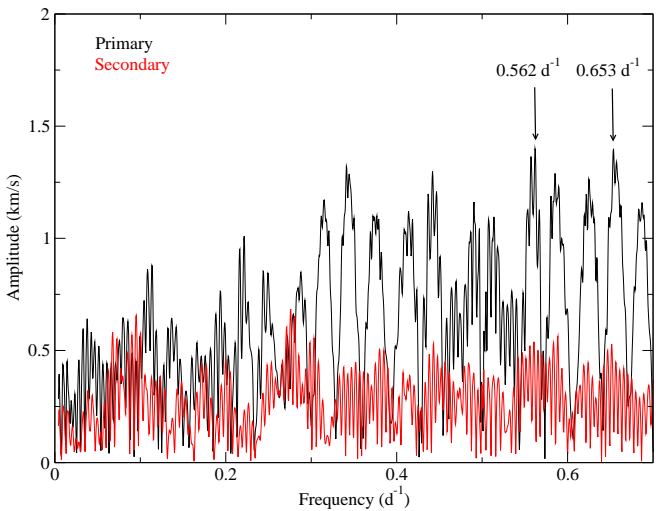}
\caption{Amplitude spectrum of the radial velocity residuals for both
components, after subtraction of the best Keplerian orbit fit. The two highest
peaks for the primary (indicated) correspond to frequencies detected in the
\textit{Kepler} light curve.}
\label{ampspec-rv-res}
\end{figure}

To study the pulsation signal in detail, we performed a complete frequency
analysis using an iterative prewhitening procedure. The Lomb-Scargle
periodogram was used in combination with false-alarm probabilities to detect the
significant frequencies present in the light curve. Prewhitening of the
frequencies was performed using linear least-squares fitting with non-linear
refinement. In total, hundreds of formally significant frequencies were
detected. We stress here that formal significance does not imply that a
frequency has a physical interpretation, e.g. in the sense of all being
independent pulsation modes. In fact, most of them are not independent at all;
many combination
frequencies and harmonics are present and many peaks arise from the fact that
we are performing Fourier analysis on a signal that is not strictly periodic.
Table \ref{freqs} lists the 50
most significant frequencies, together with their amplitude, and estimated S/N.
The noise level used to calculate the S/N was estimated from the average
amplitude in the frequency range between 20 and 24 {\cd}. The last column
indicates possible combination frequencies and harmonics. The lowest order
combination is always listed, but for some frequencies, these can be
written equivalently as a different higher order combination of more dominant
frequencies (they are listed in brackets). A complete list of frequencies
is available online in electronic format at the CDS \footnote{Centre de
Donn\'{e}es astronomiques de Strasbourg,
  http://cdsweb.u-strasbg.fr/}.
Clearly, many frequencies can
be explained by low-order linear combinations of the three most dominant
frequencies. However, given the amplitude modulation,
these dominant frequencies are likely not independent (corresponding to
different
pulsation modes). Instead, they are probably combination frequencies of the
real
pulsation frequency and (harmonics of) the modulation frequency $f_\mathrm{mod}$
$\sim$ 0.010 {\cd}. 
Looking back at the RV data for the pulsating component,
we expect the larger scatter in comparison to the secondary to be caused by the
pulsations. After subtraction of the Keplerian orbit fit, we performed
frequency analysis on the residuals, and found 2 significant
peaks corresponding to frequencies detected in the \textit{Kepler} data
(f1 and f7 in Table \ref{freqs}). The amplitude spectra of the RV residuals for
both components are shown in Fig. \ref{ampspec-rv-res}. 

From spectroscopy, we found v sin i = 14.2 $\pm$ 1.5 km/s for the pulsating
primary component and v sin i = 8.4 $\pm$ 1.5 km/s for the secondary. Assuming that the rotation axes are perpendicular to the orbital plane, and using the orbital and stellar
parameters listed in Table \ref{final-par}, this corresponds to a rotation
period of about 7.5 $\pm$ 1.3 d for the primary and 8.8 $\pm$ 1.9 d for the
secondary. These values suggest super-synchronous rotation, but on their own do
not provide sufficient evidence, given the uncertainties (1$\sigma$), especially
not for the secondary.

\begin{figure}
 \centering
   \includegraphics[width=14cm,angle=270,scale=0.5]{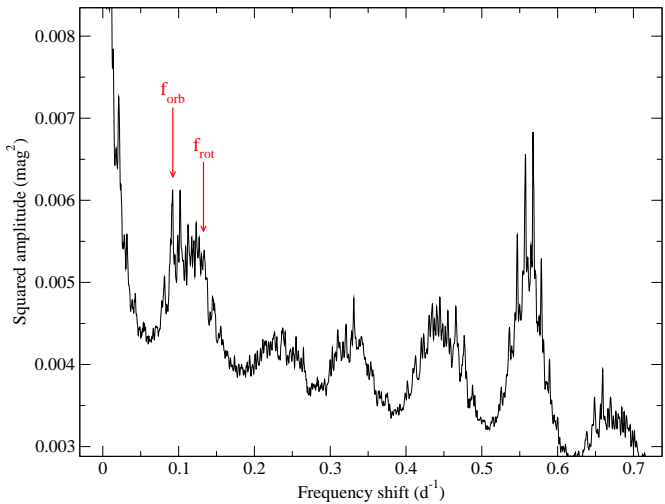}
\caption{Autocorrelation of the amplitude spectrum, after removal of the binary
model.}
\label{autocorrelation}
\end{figure}

We analysed the pulsation signal to
detect signs of rotational splitting of the pulsation frequencies in two
different ways:  by computing the autocorrelation function of the
amplitude spectrum and by using the list of detected significant frequencies.
The autocorrelation function between 0 and 0.7 {\cd} is
shown in Fig. \ref{autocorrelation}. Clearly, the amplitude spectrum is
self-similar for many different frequency shifts, as is evidenced by the
complex structure of the autocorrelation function. Although the
autocorrelation function is dominated by the highest peaks in the
amplitude spectrum and their higher harmonics, we can relate some smaller peaks
to the properties of the system. For example, the peaks indicated with
the red arrows correspond to the orbital frequency and the rotational frequency of the primary as derived from spectroscopy (at 0.0926 and 0.1333 {\cd} respectively). 

Next, we searched the list of significant frequencies for any possible frequency
differences occurring several times (given the frequency resolution). This way,
splittings of lower amplitude peaks can be detected more easily, which is not the
case when using the autocorrelation function. Since the number of significant
frequencies is so large, several frequency differences occur multiple times
purely by chance (this was checked by using a list of randomly generated
frequencies), so one must be careful when interpreting the results
\citep[see e.g.][]{Papics:2012}.
In our search for splittings, we used frequency lists of different lengths,
ranging from the first 50 to a maximum of several hundred significant
frequencies (down to a S/N of 10). We used a rather strict cutoff value of
0.0001 {\cd} to accept frequency differences
as being equal ( 0.1/T, with T the total time span of the light curve). Next, we ordered all possible frequency differences in increasing
order of occurrence. This always resulted in the same differences showing up in
the top of the lists. Table \ref{freqdiffs} shows the most abundant differences detected in a conservative list of `only' 400 frequencies (down to a S/N of about 50). 
Clearly, many of these differences are related to the
amplitude modulation in the light curve (as discussed above), with values around
0.010 {\cd} (suspected modulation frequency) and 0.020 {\cd} (twice the
modulation frequency) occurring often. Values around 0.567 {\cd} are related to
the presence of higher harmonics of the pulsation frequencies. 
In the autocorrelation function, clear peaks are present around the orbital
frequency, while the orbital frequency itself only
shows up in the list of frequency differences when going down to S/N values
around 10. Most likely, this is caused by small residuals of the binary signal
(eclipses) in the light curve. 
We also find values very close to the rotation
frequency from spectroscopy and values in between the rotation frequency and the orbital frequency. They are also visible as the group of peaks in the
autocorrelation function around 0.1 {\cd}. We interpret these as the result of
rotational splitting. The values are compatible with the spectroscopic results
obtained for the primary, and hence provide stronger evidence for
super-synchronous rotation. We cannot unambiguously identify one unique
rotational splitting, but rather a range of possible values. The difference in
measured splitting values across the spectrum suggests non-rigid rotation in
the interior of the primary \citep[see e. g.][]{Aerts:2003, Aerts:book,Dziembowski:2008}.

In Fig. \ref{as-puls-global}, the structures in the amplitude spectrum caused by
rotational splitting are indicated by means of red arrows. Clearly, the pattern
is repeated for the higher harmonics of the pulsation frequencies. 

Finally, the second-most dominant peak in
the amplitude spectrum almost coincides with the sixth harmonic of the orbital
frequency. This suggests tidally affected
pulsation, although this is not expected for non-eccentric systems.
Moreover, the frequencies do not match perfectly, and the second-most dominant
peak is likely not an independent pulsation mode, but caused by the amplitude
modulation of the main oscillation frequency.

\begin{table*}
\center
\caption{Dominant fifty frequencies with their amplitude and S/N, detected in the
\textit{Kepler} light curve with the binary model removed, using the
prewhitening technique as explained in the text.}
 \begin{tabular}{c|c|c|c|c}
Number&Frequency (d$^{-1}$) & Amplitude (mmag)& S/N&Combination\\
\hline
 1       &0.5673 &17.487&720.1&-\\
 2       &0.5575 &13.482&656.0&-\\
 3       &0.5685 &8.378 &460.2&-\\
 4       &1.1250 &7.468 &450.2&f1+f2\\
 5       &0.5467 &6.559 &400.4&2f2-f3\\
 6       &1.1359 &5.462 &355.8&f1+f3\\
 7       &0.6594 &4.704 &323.4&-\\
 8       &1.1141 &4.466 &311.7&f1+f5 (f1+2f2-f3)\\
 9       &0.5563 &4.124 &296.7&f4-f3 (f1+f2-f3)\\
 10      &0.4348 &3.192 &237.7&-\\
 11      &0.4403 &3.288 &246.5&-\\
 12      &1.6936 &2.842 &215.5&f3+f4 (f1+f2+f3)\\
 13      &0.5666 &2.547 &195.9&-\\
 14      &0.8985 &2.286 &178.8&-\\
 15      &1.1041 &2.123 &168.1&f2+f5 (3f2-f3)\\
 16      &0.3029 &2.080 &164.6&2f10-f13\\
 17      &0.4483 &2.014 &161.4&-\\
 18      &1.6925 &2.009 &163.2&f1+f4 (2f1+f2)\\
 19      &1.6716 &1.969 &159.8&f4+f5 (f1+3f2-f3)\\
 20      &0.5357 &1.963 &156.4&2f5-f2 (3f2-2f3)\\
 21      &0.2365 &1.864 &150.7&2f1-f14\\
 22      &1.1258 &1.847 &150.2&f2+f3\\
 23      &0.4777 &1.831 &150.8&2f3-f7\\
 24      &0.4337 &1.780 &148.4&2f5-f7 \\
 25      &0.5784 &1.770 &147.7&f4-f5 (f1-f2+f3)\\
 26      &1.2268 &1.766 &149.6&f1+f7\\
 27      &1.6816 &1.738 &146.2&2f1+f5 (2f1+2f2-f3)\\
 28      &0.4656 &1.712 &145.9&f4-f7 (f1+f2-f7)\\
 29      &0.1272 &1.729 &149.5&f1-f11\\
 30      &1.0023 &1.643 &142.2&f1+f10\\
 31      &0.9979 &1.667 &146.7&f2+f11\\
 32      &0.5582 &1.612 &143.8&f22-f1 (f2+f3-f1)\\
 33      &0.6583 &1.539 &138.1&2f5-f10\\
 34      &1.1346 &1.514 &135.7&2f1\\
 35      &0.2477 &1.456 &131.7&2f7-2f20\\
 36      &0.9804 &1.398 &127.6&f5+f24 (3f5-f7)\\
 37      &1.2167 &1.399 &127.5&f2+f7\\
 38      &1.7031 &1.388 &126.5&f1+f3\\
 39      &0.4115 &1.363 &124.1&f7-f35 (2f20-f7)\\
 40      &2.2609 &1.326 &122.2&f4+f6 (2f1+f2+f3)\\
 41      &0.5257 &1.308 &120.3&2f5-f1\\
 42      &0.5458 &1.267 &116.8&f8-f3 (f1+2f2-2f3)\\
 43      &0.4200 &1.272 &117.2&2f7-f14\\
 44      &0.0909 &1.240 &113.9&f7-f3\\
 45      &0.5677 &1.220 &112.9&f1\\
 46      &1.7043 &1.209 &113.8&f3+f6 (f1+2f3)\\
 47      &0.6720 &1.201 &114.5&2f9-f11\\
 48      &1.1368 &1.194 &114.4&2f3\\
 49      &1.1470 &1.160 &113.0&2f6-f4 (f1-f2+2f3)\\
 50      &0.2579 &1.159 &113.3&f4-2f24\\
\hline

\end{tabular}
\tablefoot{Typical uncertainties on the frequency values are
$\sim$ $10^{-3}$ d$^{-1}$ (using the Rayleigh criterion), and $\sim 5 \times
10^{-3}$ mmag for the amplitudes. The last column indicates possible
combination frequencies and harmonics.}
\label{freqs}
\end{table*}

\begin{table*}
\center
\caption{Frequency differences and corresponding periods (in order of decreasing occurrence), detected in the list of the first 400 frequencies obtained using iterative prewhitening.}
 \begin{tabular}{c|c|c}
Frequency difference (d$^{-1}$)&  Period (d)&Remarks\\
\hline
       0.5674  &      1.7624 & harmonics of pulsation frequencies\\ 
       0.0108  &    92.5926 & amplitude modulation\\
       0.5675  &    1.7621   & harmonics of pulsation frequencies\\
       0.1244   &     8.0386  & rotational splitting? \\ 
       0.5575   &     1.7937 &  harmonics of pulsation frequencies  \\
       0.1133   &     8.8261  &  rotational splitting? \\
       0.5685   &     1.7590   & harmonics of pulsation frequencies  \\ 
       0.5672   &     1.7630  & harmonics of pulsation frequencies \\
       0.1010   &     9.9010   & rotational splitting? \\
       0.0210  &     47.6190  &  amplitude modulation\\ 
       0.5577  &      1.7931 & harmonics of pulsation frequencies  \\
       0.3132  &      3.1928  &  \\
       0.2492  &      4.0128   &  \\ 
       0.0110  &     90.9091  &  amplitude modulation \\
       0.0284  &     35.2113   &  \\
       0.2267  &      4.4111    &  \\ 
       0.3318  &      3.0139   &  \\
       0.0109  &     91.7431    & amplitude modulation \\
       0.0070  &    142.8571  & \\ 
       0.3353  &      2.9824 &  \\
       ...         &     ... &\\
\hline                       
\end{tabular}
\label{freqdiffs}
\end{table*}

\section{Conclusions}

We have obtained accurate system parameters and astrophysical
properties for KIC 11285625, a double-lined eclipsing binary
system with a $\gamma$ Dor pulsator discovered by the \textit{Kepler} space
mission. The excellent \textit{Kepler} data with a total time span of almost
1000 days have been analysed together with high resolution HERMES spectra.
The individual composite spectra could not be used to derive fundamental
parameters such as
{\teff} and {\logg}, given their insufficient S/N. This was achieved after using
the spectral disentangling technique for both components. 

An iterative automated method was developed to separate the orbital variability
in the \textit{Kepler} light curve from the variability due to the pulsations of
the primary. The fact that the orbital frequency and its overtones are located 
in the same frequency range as the pulsation frequencies, made a simple
separation technique (such as a filter in the frequency domain) insufficient.
We plan to develop this technique further and apply it to other binary systems
in the \textit{Kepler} database.

After removal of the best binary model, we studied the residual pulsation
signal in detail, and found indications for rotational splitting of the pulsation
frequencies, compatible with super-synchronous and non-rigid internal rotation.
A detailed asteroseismic analysis of the $\gamma$ Dor pulsator and comparison
with theoretical models can now be attempted on the basis of this observational work, which
constitutes an excellent starting point for stellar modelling of a \GD\
star. A concrete interpretation of the detected amplitude modulation must await a much longer \textit{ Kepler} light curve, given the relatively long modulation period.

\begin{acknowledgements}
The research leading to these results has received funding from the European
Research Council under the European Community's Seventh Framework Programme
(FP7/2007--2013)/ERC grant agreement n$^\circ$227224 (PROSPERITY), from the
Research Council of K.U.Leuven (GOA/2008/04), and from the Belgian federal
science
policy office (C90309: CoRoT Data Exploitation); A. Tkachenko and P. Degroote
are postdoctoral fellows of the Fund for Scientific Research (FWO), Flanders,
Belgium. 
Funding for the \textit{Kepler} Discovery mission is provided by NASA's Science
Mission Directorate.
Some of the data presented in this paper were obtained from the
Multimission Archive at the Space Telescope Science Institute (MAST). STScI is
operated by the Association of Universities for Research in Astronomy, Inc.,
under NASA contract NAS5-26555. Support for MAST for non-HST data is provided by
the NASA Office of Space Science via grant NNX09AF08G and by other grants and
contracts.
This research has made use of the SIMBAD database,
operated at CDS, Strasbourg, France.
We would like to express our special thanks to the numerous people who helped
make the \textit{Kepler} mission possible.
\end{acknowledgements}

\bibliographystyle{aa}
\bibliography{references}

\begin{thebibliography}{34}
\expandafter\ifx\csname natexlab\endcsname\relax\def\natexlab#1{#1}\fi

\bibitem[{{Aerts} {et~al.}(2010){Aerts}, {Christensen-Dalsgaard}, \&
  {Kurtz}}]{Aerts:book}
{Aerts}, C., {Christensen-Dalsgaard}, J., \& {Kurtz}, D.~W. 2010,
  {Asteroseismology} (Springer)

\bibitem[{{Aerts} {et~al.}(2003){Aerts}, {Thoul}, {Daszy{\'n}ska}, {Scuflaire},
  {Waelkens}, {Dupret}, {Niemczura}, \& {Noels}}]{Aerts:2003}
{Aerts}, C., {Thoul}, A., {Daszy{\'n}ska}, J., {et~al.} 2003, Science, 300,
  1926

\bibitem[{{Balona}(2012)}]{Balona:2012}
{Balona}, L.~A. 2012, \mnras, 422, 1092

\bibitem[{{Benk{\' o}} {et~al.}(2011){Benk{\' o}}, {Szab{\'o}}, \&
  {Papar{\'o}}}]{Benko:2011}
{Benk{\' o}}, J.~M., {Szab{\'o}}, R., \& {Papar{\'o}}, M. 2011, \mnras, 417,
  974

\bibitem[{{Borucki} {et~al.}(2010){Borucki}, {Koch}, {Basri}, {Batalha},
  {Brown}, {Caldwell}, {Caldwell}, {Christensen-Dalsgaard}, {Cochran},
  {DeVore}, {Dunham}, {Dupree}, {Gautier}, {Geary}, {Gilliland}, {Gould},
  {Howell}, {Jenkins}, {Kondo}, {Latham}, {Marcy}, {Meibom}, {Kjeldsen},
  {Lissauer}, {Monet}, {Morrison}, {Sasselov}, {Tarter}, {Boss}, {Brownlee},
  {Owen}, {Buzasi}, {Charbonneau}, {Doyle}, {Fortney}, {Ford}, {Holman},
  {Seager}, {Steffen}, {Welsh}, {Rowe}, {Anderson}, {Buchhave}, {Ciardi},
  {Walkowicz}, {Sherry}, {Horch}, {Isaacson}, {Everett}, {Fischer}, {Torres},
  {Johnson}, {Endl}, {MacQueen}, {Bryson}, {Dotson}, {Haas}, {Kolodziejczak},
  {Van Cleve}, {Chandrasekaran}, {Twicken}, {Quintana}, {Clarke}, {Allen},
  {Li}, {Wu}, {Tenenbaum}, {Verner}, {Bruhweiler}, {Barnes}, \&
  {Prsa}}]{Borucki-Kepler}
{Borucki}, W.~J., {Koch}, D., {Basri}, G., {et~al.} 2010, Science, 327, 977

\bibitem[{{Breger} {et~al.}(2011){Breger}, {Balona}, {Lenz}, {Hollek}, {Kurtz},
  {Catanzaro}, {Marconi}, {Pamyatnykh}, {Smalley}, {Su{\'a}rez}, {Szabo},
  {Uytterhoeven}, {Ripepi}, {Christensen-Dalsgaard}, {Kjeldsen}, {Fanelli},
  {Ibrahim}, \& {Uddin}}]{Breger:2011}
{Breger}, M., {Balona}, L., {Lenz}, P., {et~al.} 2011, \mnras, 414, 1721

\bibitem[{{Debosscher} {et~al.}(2011){Debosscher}, {Blomme}, {Aerts}, \& {De
  Ridder}}]{Debosscher:2011}
{Debosscher}, J., {Blomme}, J., {Aerts}, C., \& {De Ridder}, J. 2011, \aap,
  529, A89

\bibitem[{{Degroote} {et~al.}(2009){Degroote}, {Briquet}, {Catala},
  {Uytterhoeven}, {Lefever}, {Morel}, {Aerts}, {Carrier}, {Auvergne}, {Baglin},
  \& {Michel}}]{Degroote:2009}
{Degroote}, P., {Briquet}, M., {Catala}, C., {et~al.} 2009, \aap, 506, 111

\bibitem[{{Dziembowski} \& {Pamyatnykh}(2008)}]{Dziembowski:2008}
{Dziembowski}, W.~A. \& {Pamyatnykh}, A.~A. 2008, \mnras, 385, 2061

\bibitem[{{Etzel}(1981)}]{Etzel:EBOP}
{Etzel}, P.~B. 1981, in Photometric and Spectroscopic Binary Systems, ed.
  {E.~B.~Carling \& Z.~Kopal}, 111

\bibitem[{{Grevesse} {et~al.}(2007){Grevesse}, {Asplund}, \&
  {Sauval}}]{Grevesse2007}
{Grevesse}, N., {Asplund}, M., \& {Sauval}, A.~J. 2007, \ssr, 130, 105

\bibitem[{{Hadrava}(1995)}]{Hadrava:1995}
{Hadrava}, P. 1995, \aaps, 114, 393

\bibitem[{{Handler} \& {Shobbrook}(2002)}]{Handler:DSCUT}
{Handler}, G. \& {Shobbrook}, R.~R. 2002, \mnras, 333, 251

\bibitem[{{Hensberge} {et~al.}(2000){Hensberge}, {Pavlovski}, \&
  {Verschueren}}]{Hensberge:2000}
{Hensberge}, H., {Pavlovski}, K., \& {Verschueren}, W. 2000, \aap, 358, 553

\bibitem[{{Ilijic} {et~al.}(2004){Ilijic}, {Hensberge}, {Pavlovski}, \&
  {Freyhammer}}]{Ilijic:2004}
{Ilijic}, S., {Hensberge}, H., {Pavlovski}, K., \& {Freyhammer}, L.~M. 2004, in
  Astronomical Society of the Pacific Conference Series, Vol. 318,
  Spectroscopically and Spatially Resolving the Components of the Close Binary
  Stars, ed. R.~W. {Hilditch}, H.~{Hensberge}, \& K.~{Pavlovski}, 111--113

\bibitem[{{Kupka} {et~al.}(2000){Kupka}, {Ryabchikova}, {Piskunov}, {Stempels},
  \& {Weiss}}]{Kupka2000}
{Kupka}, F.~G., {Ryabchikova}, T.~A., {Piskunov}, N.~E., {Stempels}, H.~C., \&
  {Weiss}, W.~W. 2000, Baltic Astronomy, 9, 590

\bibitem[{{Lehmann} {et~al.}(2011){Lehmann}, {Tkachenko}, {Semaan},
  {Guti{\'e}rrez-Soto}, {Smalley}, {Briquet}, {Shulyak}, {Tsymbal}, \& {De
  Cat}}]{Lehmann2011}
{Lehmann}, H., {Tkachenko}, A., {Semaan}, T., {et~al.} 2011, \aap, 526, A124

\bibitem[{{Maceroni} {et~al.}(2013){Maceroni}, {Montalb{\'a}n}, {Gandolfi},
  {Pavlovski}, \& {Rainer}}]{Maceroni:2013}
{Maceroni}, C., {Montalb{\'a}n}, J., {Gandolfi}, D., {Pavlovski}, K., \&
  {Rainer}, M. 2013, \aap, in press

\bibitem[{{Maceroni} {et~al.}(2009){Maceroni}, {Montalb{\'a}n}, {Michel},
  {Harmanec}, {Prsa}, {Briquet}, {Niemczura}, {Morel}, {Ladjal}, {Auvergne},
  {Baglin}, {Baudin}, {Catala}, {Samadi}, \& {Aerts}}]{Maceroni:2009}
{Maceroni}, C., {Montalb{\'a}n}, J., {Michel}, E., {et~al.} 2009, \aap, 508,
  1375

\bibitem[{{P{\'a}pics}(2012)}]{Papics:2012}
{P{\'a}pics}, P.~I. 2012, Astronomische Nachrichten, 333, 1053

\bibitem[{{Pavlovski} \& {Hensberge}(2005)}]{Pavlovski:2005}
{Pavlovski}, K. \& {Hensberge}, H. 2005, \aap, 439, 309

\bibitem[{{Popper} \& {Etzel}(1981)}]{Popper:EBOP}
{Popper}, D.~M. \& {Etzel}, P.~B. 1981, \aj, 86, 102

\bibitem[{{Poretti} {et~al.}(2011){Poretti}, {Rainer}, {Weiss}, {Bogn{\'a}r},
  {Moya}, {Niemczura}, {Su{\'a}rez}, {Auvergne}, {Baglin}, {Baudin}, {Benk{\H
  o}}, {Debosscher}, {Garrido}, {Mantegazza}, \& {Papar{\'o}}}]{Poretti:2011}
{Poretti}, E., {Rainer}, M., {Weiss}, W.~W., {et~al.} 2011, \aap, 528, A147

\bibitem[{{Pr{\v s}a} {et~al.}(2011){Pr{\v s}a}, {Batalha}, {Slawson}, {Doyle},
  {Welsh}, {Orosz}, {Seager}, {Rucker}, {Mjaseth}, {Engle}, {Conroy},
  {Jenkins}, {Caldwell}, {Koch}, \& {Borucki}}]{Prsa:2011}
{Pr{\v s}a}, A., {Batalha}, N., {Slawson}, R.~W., {et~al.} 2011, \aj, 141, 83

\bibitem[{{Raskin} {et~al.}(2011){Raskin}, {van Winckel}, {Hensberge},
  {Jorissen}, {Lehmann}, {Waelkens}, {Avila}, {de Cuyper}, {Degroote},
  {Dubosson}, {Dumortier}, {Fr{\'e}mat}, {Laux}, {Michaud}, {Morren}, {Perez
  Padilla}, {Pessemier}, {Prins}, {Smolders}, {van Eck}, \& {Winkler}}]{HERMES}
{Raskin}, G., {van Winckel}, H., {Hensberge}, H., {et~al.} 2011, \aap, 526, A69

\bibitem[{{Rodr{\'{\i}}guez} \& {Breger}(2001)}]{Rodriguez:2001}
{Rodr{\'{\i}}guez}, E. \& {Breger}, M. 2001, \aap, 366, 178

\bibitem[{{Shulyak} {et~al.}(2004){Shulyak}, {Tsymbal}, {Ryabchikova},
  {St{\"u}tz}, \& {Weiss}}]{Shulyak2004}
{Shulyak}, D., {Tsymbal}, V., {Ryabchikova}, T., {St{\"u}tz}, C., \& {Weiss},
  W.~W. 2004, \aap, 428, 993

\bibitem[{{Simon} \& {Sturm}(1994)}]{Simon:1994}
{Simon}, K.~P. \& {Sturm}, E. 1994, \aap, 281, 286

\bibitem[{{Southworth} {et~al.}(2004{\natexlab{a}}){Southworth}, {Maxted}, \&
  {Smalley}}]{Southworth:JKTEBOP1}
{Southworth}, J., {Maxted}, P.~F.~L., \& {Smalley}, B. 2004{\natexlab{a}},
  \mnras, 351, 1277

\bibitem[{{Southworth} {et~al.}(2004{\natexlab{b}}){Southworth}, {Zucker},
  {Maxted}, \& {Smalley}}]{Southworth:JKTEBOP2}
{Southworth}, J., {Zucker}, S., {Maxted}, P.~F.~L., \& {Smalley}, B.
  2004{\natexlab{b}}, \mnras, 355, 986

\bibitem[{{Tkachenko} {et~al.}(2013){Tkachenko}, {Aerts}, {Yakushechkin},
  {Debosscher}, {Degroote}, {Bloemen}, {Papics}, {de Vries}, {Lombaert},
  {Hrudkova}, {Fremat}, {Raskin}, \& {Van Winckel}}]{Tkachenko:2013}
{Tkachenko}, A., {Aerts}, C., {Yakushechkin}, A., {et~al.} 2013, ArXiv
  e-prints: 1305.6722

\bibitem[{{Tkachenko} {et~al.}(2012){Tkachenko}, {Lehmann}, {Smalley},
  {Debosscher}, \& {Aerts}}]{Tkachenko2012}
{Tkachenko}, A., {Lehmann}, H., {Smalley}, B., {Debosscher}, J., \& {Aerts}, C.
  2012, \mnras, 422, 2960

\bibitem[{{Tsymbal}(1996)}]{Tsymbal1996}
{Tsymbal}, V. 1996, in Astronomical Society of the Pacific Conference Series,
  Vol. 108, M.A.S.S., Model Atmospheres and Spectrum Synthesis, ed. S.~J.
  {Adelman}, F.~{Kupka}, \& W.~W. {Weiss}, 198

\bibitem[{{Welsh} {et~al.}(2011){Welsh}, {Orosz}, {Aerts}, {Brown},
  {Brugamyer}, {Cochran}, {Gilliland}, {Guzik}, {Kurtz}, {Latham}, {Marcy},
  {Quinn}, {Zima}, {Allen}, {Batalha}, {Bryson}, {Buchhave}, {Caldwell},
  {Gautier}, {Howell}, {Kinemuchi}, {Ibrahim}, {Isaacson}, {Jenkins}, {Prsa},
  {Still}, {Street}, {Wohler}, {Koch}, \& {Borucki}}]{Welsh:2011}
{Welsh}, W.~F., {Orosz}, J.~A., {Aerts}, C., {et~al.} 2011, \apjs, 197, 4

\end{thebibliography}

\end{document}